# Trust and Uncertainty in Strategic Interaction: Behavioural and Physiological Evidence from the Centipede Game

*Dhiraj Jagadale (Dhiraj.jagadale@researchi.iiit.ac.in), Kavita Vemuri\* (kvemuri@iiit.ac.in)*

*International Institute of Information Technology, Hyderabad*

*C.R.Rao Road, Gachibowli, Hyderabad 500032*

(*[\*kvemuri@iiit.ac.in](*kvemuri@iiit.ac.in)*)

Abstract

Mutual trust is a key determinant of decision-making in economic interactions, yet actual behavior often diverges from equilibrium predictions. This study investigates how emotional arousal, indexed by skin conductance responses (SCR), relates to trust behavior in a modified centipede game. To examine the impact of uncertainty, the game incorporated both fixed and random termination conditions. SCRs were recorded alongside self-reported measures of mutual and general trust and individual risk-taking propensity.

Phasic SCRs were significantly higher under random termination, particularly following the opponent's "take" actions, indicating increased emotional arousal under uncertainty. Mutual trust scores correlated positively with risk propensity but not with general trust. Behaviorally, higher mutual trust was associated with extended cooperative play, but only in the fixed-turn condition.

These findings suggest that physiological arousal reflects emotional engagement in trust-related decisions and that uncertainty amplifies both arousal and strategic caution. Mutual trust appears context-dependent, shaped by emotional and physiological states that influence deviations from equilibrium behavior.

## 1.0 Introduction

Emotions play a critical role in regulating decision-making, particularly in social and financial contexts, influencing both individual and collective choices (Naqvi et al., 2006; Grecucci & Sanfey, 2013; Grecucci et al., 2019; Heilman et al., 2016; van Dijk & De Dreu, 2021). For example, heightened anger or frustration is linked to increased rejection of offers (Pillutla & Murnighan, 1996; Sanfey et al., 2003), while sadness or negative mood has been shown to lead to lower acceptance of fair offers (Harle & Sanfey, 2007; Forgas & Tan, 2013), particularly in Ultimatum Game (UG) experiments. Further research demonstrates that emotion regulation—such as reappraisal or distancing from the emotional context—can modulate emotional responses and subsequent decision-making (Grecucci, et al. 2015; Grecucci et al., 2019). More recent work by Grecucci et al. (2020) highlights how emotion regulation strategies can affect choices in games like UG and the Dictator Game, showing how emotional responses can influence fairness-based decisions. Lizotte et al. (2021) proposed a unified process-level account to understand the empirical findings from games such as UG and the Prisoner's Dilemma, suggesting that emotion-driven processes play a significant role in shaping decision outcomes.

Studies have shown that SCR is linked to the somatic marker hypothesis, which posits that emotional signals, such as changes in skin conductance, guide decision-making without conscious awareness (Bechara et al., 1997). A form of EDA, it is a reliable measure of emotional arousal, with changes in skin conductivity reflecting increased sympathetic nervous system activity (Bradley et al., 2008). SCR is triggered by a burst of activity in the sweat glands, which is often associated with emotional responses to stimuli, such as fear, anxiety, or cognitive load (Lang & Davis, 2006; Canli & Lesch, 2007). In decision-making studies, SCR serves as a proxy for emotional arousal, providing insight into the emotional processes that underlie choices, particularly in uncertain or risky situations (Bechara et al., 2005; Figner & Murphy, 2011). Functional neuroimaging and SCR research have demonstrated that both cognitive and emotional processes can influence the skin conductance response, revealing how the brain integrates emotional signals to make decisions (Critchley, 2002).

In this study, we aim to understand the dynamics between mutual trust, risk-taking, and decision-making in the Centipede Game by correlating participants' decisions with their SCR at two key events: the decision to "take" or "pass." The findings will deepen our understanding of how emotional arousal, captured through SCR, influences decision-making in trust-based economic games. By examining the Centipede Game, this study explores the complex interplay between mutual trust, risk-taking, and physiological responses, offering new insights into the role of emotions in decision-making under uncertainty. In doing so, it contributes to the broader literature on trust, cooperation, and emotional processes in human interactions.

The hypotheses are as follows:

H0: In *known_rounds* condition, the number of cooperative turns increases as a function of trust

H1: In the *known_rounds* condition, where players know the number of rounds, mutual trust will increase the number of cooperative turns, and SCR will be lower due to reduced anxiety.

H2: In the *random_rounds* condition, where the game ends unpredictably, trust will have less impact on cooperation, and SCR will be higher due to increased uncertainty and anxiety.

H3: Self decision process of either a 'take' or 'pass' will have higher anxiety level

## 1.1 Literature Review

### 1.1.1 Electrodermal Activity (EDA) and Emotional Arousal in Decision-Making

Electrodermal Activity (EDA), which includes measures like Galvanic Skin Response (GSR), is a key tool for studying the physiological basis of emotional arousal and its impact on decision-making (Benedek & Kaernbach, 2011; Figner & Murphy, 2011). EDA reflects changes in the skin's electrical conductivity, which is influenced by the activity of sweat glands—a process triggered by sympathetic nervous system arousal (Kennedy et al., 1994; Riedl et al., 1998). GSR, as a measure of emotional arousal, has been widely used in research to assess the affective processes that underlie decisions, particularly in contexts involving risk and uncertainty (Bechara et al., 2005; Dunn et al., 2006; Van't Wout et al., 2006; Dawson et al., 2011).

Skin Conductance Response (SCR) has proven to be a reliable marker of emotional engagement, distinguishing between positive and negative emotions, and even specific emotional states such as fear, anxiety, and anger (Canli & Lesch, 2007; Bradley et al., 2008). Studies on decision-making have shown that SCR is sensitive to emotionally charged decisions, such as those encountered in economic games like the Ultimatum Game (Sarlo et al., 2012; Hewig et al., 2011), the Trust Game (Ibáñez et al., 2016), and the Balloon Analog Risk Task (Hüpen et al., 2019). These findings underscore the importance of physiological responses in understanding the emotional underpinnings of decision-making.

In particular, SCR has been linked to the concept of somatic markers—unconscious emotional signals that guide decision-making even when individuals are unaware of the emotional processes at play (Bechara et al., 1997). The anticipatory emotional responses reflected in SCR provide insights into the affective states that influence how people navigate risky decisions (Figner & Murphy, 2011).

1.1.2 The Centipede Game: Trust, Risk, and Emotional Responses

This study aims to extend our understanding of the relationship between emotional arousal and decision-making by examining the Centipede Game—a well-established economic game that involves sequential decision-making with cooperative and competitive elements (Rosenthal, 1981; Schotter & Sopher, 2006). In this game, players face the decision of whether to continue cooperating or "take" (defect) to secure a larger personal payoff. Cooperation in the Centipede Game requires trust, as players must rely on their opponents to make mutually beneficial decisions.

Trust, however, is inherently fraught with uncertainty. As Lascaux (2008) suggests, the unknown behavior of the opponent introduces anxiety, which can affect decision-making. This anxiety is often captured in skin conductance responses, providing a physiological measure of the emotional processes at play. When players face the decision to "pass" (cooperate), anxiety may arise due to the uncertainty of the opponent's response. This uncertainty is especially pronounced in the dyadic nature of the game, where each player's decision is contingent on the other's behavior. As players engage in the game, the physiological responses recorded through SCR offer a window into the anticipatory emotions that guide their decisions.

Our study specifically explores the correlation between SCR and cooperative decisions (taking or passing) in the Centipede Game. The premise is that players experience anticipatory anxiety driven by both their own choices (whether to pass or take) and the uncertainty surrounding the opponent's moves. By examining SCR during these key moments in the game, we can gain insight into how emotions, such as trust and anxiety, modulate decision-making.

1.1.3 Trust, Risk, and Cultural Differences in Decision-Making

Trust is a fundamental component of human interaction and cooperation, especially in economic decision-making (Mayer et al., 1995). In the Centipede Game, trust and risk are closely interlinked, with players deciding whether to cooperate based on their assessment of the other player's trustworthiness. According to Dawes and Thaler (1988), individuals are more likely to cooperate when they believe that the other player is trustworthy and will reciprocate their cooperation. Trust is thus a crucial factor in determining cooperative behaviors, particularly when the outcomes of decisions depend on the actions of others.

Cultural factors can also influence trust-related decision-making. Studies suggest that players from collectivist cultures, such as India, may be more cooperative in economic games, although this hypothesis remains underexplored in high-incentive, high-uncertainty contexts (Camerer, 2003; Ghosh et al., 2019). This study contributes to the literature by examining how trust, risk, and uncertainty interact in the Centipede Game, particularly in a diverse population where family ties or social connections may not be present to influence decision-making.

## 2.0 METHODOLOGY

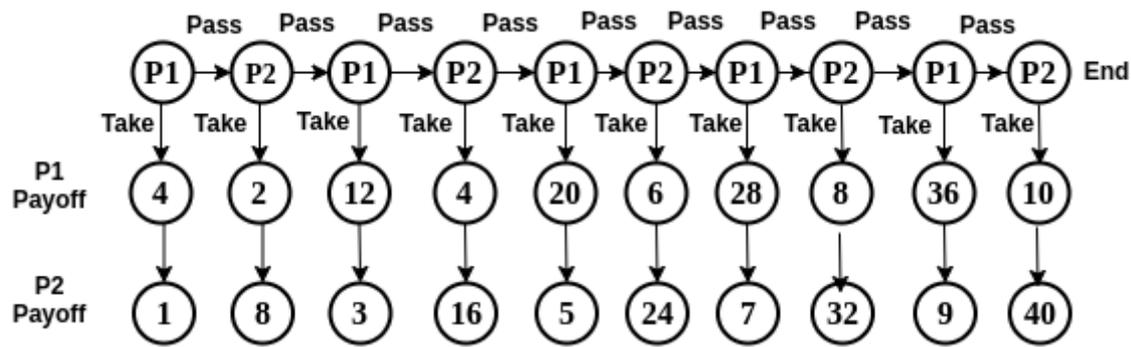

Fig.1 Normal Representation of 2-player Centipede Game. P1 is player one, and P2 is player 2. The payoff in the event of a *pass* is lower for the player.

### 2.1 Game setup

The Centipede Game setup with the payoffs at each *take* or *pass* is shown in Fig.1). The payoffs were designed to present multiple Nash Equilibria and proportionally to both players. For example, if P1 passes, the first move will end with a smaller payoff if P2 takes. Similarly, if P2 passes and P1 takes in the third turn, the payoff is significantly lower (3 compared to 8). To examine the role of trust (mutual/reciprocal), the payoffs considered are optimal to test for decision change and hence SCR change at each node. Players played two blocks of 10 games each, one block had the *known_rounds* and the second block had *random_rounds*. In *known_rounds*, the game would go up to 10 turns if both players continued passing the stacks. In *random_rounds*, the computer can end the game with the probability of ending increasing linearly from 0 (round 1) to 1(round 10). In the *known_rounds games,* the players were informed of the number of rounds, while in the *random_rounds games,* the current turn number and probability of ending after each round was displayed. A participant in each pair (labelled as '*player*') was only fitted with the GSR sensor. The *Player* and *Opponent* started the game alternatively so that none of them had the advantages of first mover. Participants were paid the payoff they earned from random six games each fromm *known_rounds* and *random_rounds* games. A payoff of 4 was converted to Rs 4/- (INR). Hence a *take* by P2 at turn 10 earns Rs.40. In the ten games played, a maximum of Rs 291 was earned by a player.

Players were asked to fill out a 5-point questionnaire (based on Robbins, 2019) to estimate players' mutual trust or trustworthiness. The questions were re-worded to reflect the culture-specific context for trust (questionnaire in supplementary). For the analysis, the pairwise trust score was used to group as low or high trust pairs (9 pairs were treated as high trust group and seven pairs of players were low trust group). The median score on the trust scores was considered as a threshold to group the pairs. The second set of questions on individual general trust, familiarity (2 question index), and risk-taking (based on Weber et al., 2002) was administered. The nonparametric Spearman correlation analysis was applied to look for significant associations between the parameters and SCR. The nonparametric analysis was considered, as the number of 'take' choices was not uniform across each participant. The relationship between behavioural scores and game choices is more monotonic than linear.

2.2 Participants

32 University students (24 Male, 8 Female, mean = 20.19±3.13) participated in the experiment. Sixteen pairs were randomly selected, with no gender pair distribution, and each pair was labelled as a player(P) and Opponent (O). A computerized version was designed and presented on two separate monitors. The players were aware of the identity and presence of each other but during the gameplay, neither verbal nor gestures communication was allowed. The face was obstructed by the monitors and an opaque partition between desks. Skin Conductance data was recorded by sensors (Procomp Infinity System, Thought Technology) fitted on the index and middle fingers of the player's(*player*) left hand.

2.3 Analysis of SCR data

All decisions were classified into four nodes: Player Pass, Player Take, Opponent Pass, and Opponent Take. Each event started at the point when payoffs for the round were shown on the screen, and the onsets were used to analyze the data. The SCR measurements were analyzed using Ledalab (Benedek & Kaernbach, 2010), a MATLAB-based tool. Data were filtered using bandpass filter {Cut-off - 1Hz - 5 Hz} and smoothed using the adaptive filter provided in the toolbox. Using Continuous Decomposition Analysis (Benedek & Kaernbach, 2010), skin conductance was further divided into slowly changing Tonic Skin conductance Level and fast-changing Phasic Skin conductance. Phasic skin conductance from 1-5 seconds after the onset of the event was used for further analysis.

3.0 RESULTS

The average number of rounds was calculated for the 16 players and their corresponding opponents. As illustrated in Figure 2, the mean 'take' decision occurred around the 6th round in the *known_rounds* block (Mean = 5.7, Median = 5.5, SD = 3.3), corresponding to a state in which the smaller stack contained 6 rupees and the larger stack 24 rupees (see Figure 1). In contrast, in the *random_rounds* block, the average 'take' occurred earlier, at approximately the 4th round (Mean = 3.6, Median = 3, SD = 1.8), when the smaller stack contained 4 rupees and the larger stack 16 rupees. This pattern was consistent across both player and opponent decisions, suggesting a tendency toward greater mutual trust when the total number of rounds was fixed. Notably, the *known_rounds* condition exhibited an approximately normal distribution around the median, with a large interquartile range, whereas the *random_rounds* condition displayed a more compact distribution with a smaller interquartile range. Consistent with prior research, the subgame perfect equilibrium prediction was not observed.

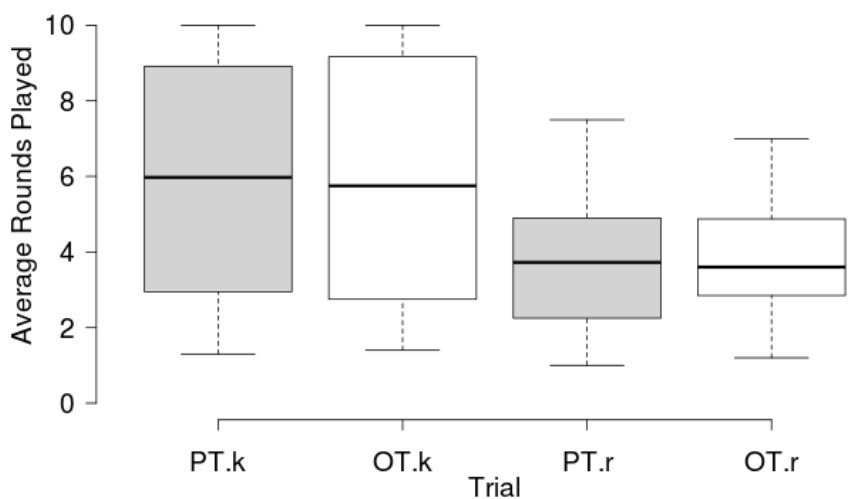

Fig. 2: Average number of rounds vs. 'Take' trials in *known_rounds* and *random_rounds* games. PT: Player Take; OT: Opponent Take; PT: Player Take; OT: Opponent Take (.k denotes *known_rounds game*, and .r is for *random_rounds* game)

As shown in Figure 3a, within the *known_rounds* condition**,** higher mutual trust was associated with a greater number of cooperative rounds before taking, and the trendline indicates that participants with higher trust scores tended to delay the take decision. This finding supports the null hypothesis (H₀), which posits that in the *known_rounds* condition, the number of cooperative turns increases as a function of trust. Conversely, in the *random_rounds* condition (Figure 3b), this relationship was not evident: participants with low trust scores showed a consistent take round across games, while those with high trust scores exhibited a slight decrease in the number of rounds. This outcome partially supports the corresponding hypothesis for the *random_rounds* condition.

3.1 Evolution of Trust in Known_rounds and Random_rounds Games

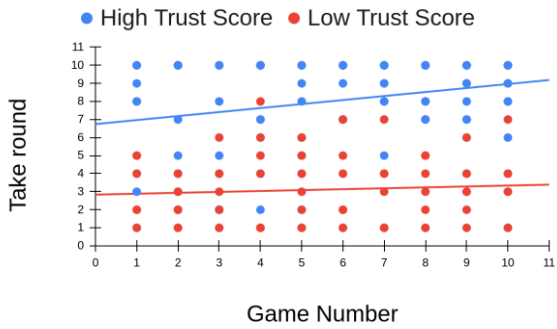 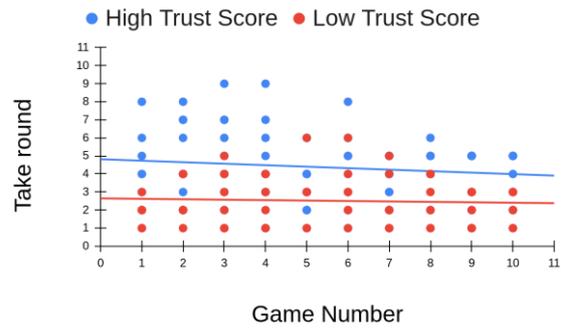

Fig. 3a                                                      Fig. 3b

Figure 3: *Take round* vs Game Number in the *known_rounds* games (Fig. 3a) and *random_rounds* games (Fig. 3b) for High Trust score and Low Trust score Pairs

Table 1: Spearman Correlation of the player's and opponent's Trust and risk score with average *Take round* in each block of games. (N: Players = 16, Opponents = 16)

| Condition | Known rounds | Random rounds | Condition | Known rounds | Random rounds |
|---|---|---|---|---|---|
| Player's Trust score vs. Player's avg. *Take round* | 0.53* | 0.31 | Player's Risk score vs. Player's avg. *Take round* | 0.27 | 0.33 |
| Opponent's Trust score vs. Opponent's avg. *Take round* | 0.58* | 0.47 | Opponent's Risk score vs. Opponent's avg. *Take round* | 0.26 | 0.29 |

\* - The correlations are significant at p < .05.

Table 1 presents the Spearman correlation coefficients between trust and risk scores and the take-round number across the two game conditions. In both *known_rounds* and *random_rounds*, the trust score was positively correlated with the take-round number, although statistical significance ($p < 0.05$) was observed only in the *known_rounds* condition. In the random_rounds, the correlation coefficient for opponent trust score was higher but did not reach significance. The risk score correlations with take-round number were positive and of similar magnitude across both conditions, though not statistically significant.

As shown in Table 2, the trust score exhibited a strong positive correlation with familiarity, as expected. However, general (dispositional) trust, which reflects an individual's overall perception of others' trustworthiness, showed no correlation with the game-specific trust score. In contrast, non-game-related risk-taking demonstrated a significant positive correlation with the trust score.

Table 2: Spearman Correlations values of Trust score, familiarity, and risk score (N = 32)

| | Familiarity with Opponent | General Trust | Risk score |
|---|---|---|---|
| With Trust Score. | 0.82* | 0.05 | 0.41* |

\* - The correlations are significant at p < .05.

3.2 Skin Conductance analysis:

The player's phasic skin conductance response (SCR) was significantly higher during the *random_rounds* blocks compared to the *known_rounds* blocks for both the player's pass and the opponent's pass/take decisions, thereby supporting hypotheses H1 and H2. The observed differences in SCR across the two game conditions indicate a differential activation of the arousal system (Boucsein, 2012). Following tests for normality, a two-tailed t-test revealed statistically significant differences ($p < 0.05$) for all gameplay decisions, with the exception of the player_take condition (Table 3). This result suggests that the anxiety associated with the decision to *take* was comparable across participants in both the *known_rounds* and *random_rounds* conditions. In contrast, the decision to *pass*—whether made by the player or the opponent—and the opponent's *take* decision elicited significantly elevated arousal levels. This pattern of results partially supports H3, suggesting that the heightened anxiety in response to the opponent's decisions reflects a measurable effect of uncertainty and diminished control over the opponent's actions.

Table 3: Average phasic SCR of the players for all trials in *known_rounds* and *random_rounds* games.

| Avg. Phasic SCR | Player Pass | Player Take | Opponent Pass | Opponent Take |
|---|---|---|---|---|
| *Known_rounds* | 0.0396 | 0.0427 | 0.0400 | 0.0389 |
| *Random_rounds* | 0.0478 | 0.0502 | 0.0476 | 0.0533 |
| T-Test (p<0.050) | 0.0195* | 0.1754 | 0.0334* | 0.0298* |

\* - two-tailed T-test was significant at p < .05.

The Spearman correlation analysis conducted to examine the relationship between SCR measurements and game trust and risk scores revealed no significant associations under the *known_rounds* condition, although the correlation coefficients were negative and relatively high in magnitude. In contrast, under the *random_rounds* condition, a strong and significant negative correlation was observed between SCR and the risk score for the player's 'take' decision (Table 4).

Table 4: Spearmann correlation of average phasic SCR of the player with trust and risk

score of the player for all trials in *random_rounds games*

| Avg. Phasic SCR | Player Pass (N = 10) | Player Take (N=8) | Opponent Pass (N = 11) | Opponent Take (N = 6) |
|---|---|---|---|---|
| Trust Score | -0.56 | 0.04 | -0.40 | -0.66 |
| Risk Score | -0.40 | -0.88* | -0.02 | -0.14 |

\* - Correlation was significant at p < .05

4.0 Discussion

The decision to trust is not a fixed choice but one shaped by ambiguity, context, and mutual interaction. In this study, we investigated how uncertainty influences trust in decision-making, particularly through the use of two distinct conditions in the centipede game: known_rounds (where the number of rounds is fixed) and *random_rounds* (where the number of rounds is uncertain). Trust, a key factor in cooperation, can fluctuate depending on the context. For instance, players may start the game with higher trust due to lower stakes or familiarity, but as the game progresses and incentives grow, trust may evolve or be challenged. Our goal was to dissect the decision-making process under uncertainty, focusing on how trust and risk-taking influence gameplay choices.

Existing research has shown a strong connection between behaviors such as risk-taking and trustworthiness in the centipede game, but these connections are often confounded by biases in self-reporting and inaccurate predictions of an opponent's strategies. As a result, understanding the reasoning behind each decision is challenging. This paper sought to clarify this process, highlighting how uncertainty shapes trust and risk-taking across different stages of the game.

We extended previous research on repeated games with fixed and random termination rules by altering the payoff structure at each turn in the centipede game. By introducing uncertainty and examining trust in this context, our study explores a nuanced but crucial aspect of human behavior: whether trust can sustain cooperation when faced with uncertainty. Mutual trust was tested at various nodes by manipulating payoffs in this experiment. Our results show a positive correlation between trust scores and the number of rounds played. High-trust pairs in the *known_rounds* condition tended to play longer, extending the game further. Conversely, low-trust pairs displayed similar behavior in both conditions, suggesting a clear link between trust propensity and risk-taking behavior in game choices.

In the *known_rounds* condition, players had perfect information, and backward induction (the strategic reasoning based on expected outcomes) led to higher trust and longer gameplay. This finding aligns with Mayer et al. (1995), who suggested that trust fosters risk-taking behavior in situations of partial certainty. However, in the *random_rounds* condition, where additional uncertainty is introduced, high-trust pairs played more conservatively, similar to low-trust pairs. This shift may be attributed to the fact that the increased uncertainty undermines mutual trust, a point supported by Engelmann et al. (2019), who found that heightened uncertainty leads to more cautious decision-making due to aversive emotions, which hinder trust.

Further analysis of backward induction in both game types revealed that players, especially those in the *known_rounds* condition, were more likely to cooperate in the earlier rounds but became more cautious as the game neared its end. This behavior mirrors findings from Cochard et al. (2004), who observed similar patterns in repeated investment games, with trust typically increasing in the early periods but decreasing near the final round.

These results also suggest that trust is not a stable trait but is sensitive to situational factors, particularly uncertainty. When certainty is high, trust facilitates cooperation; however, when uncertainty increases, competition begins to dominate, supporting Lascaux's (2008) argument that competition trumps cooperation in uncertain contexts. The influence of cultural factors on trust and cooperation was also considered. While collectivist cultures (e.g., Japan, India) may be expected to cooperate more, and individualistic cultures (e.g., the USA) to compete more, our findings did not consistently align with these cultural stereotypes, particularly in the *random_rounds* condition. Indian participants, for example, displayed more competitive behavior than expected, possibly due to the nature of the game itself, which was designed to provoke competition.

The relationship between familiarity and trust was particularly striking. Players who were familiar with their opponents had significantly higher trust scores, which aligns with social theories of trust. However, general trust measures, such as dispositional trust, did not predict dyadic trust in this context. This is consistent with Yamagishi's (2015) work, which showed that general trust does not adequately explain behavior in one-on-one interactions. Additionally, the positive correlation between self-reported risk-taking and trust propensity suggests that individuals who are more willing to take risks are also more inclined to trust their opponents, particularly in the *known_rounds* condition.

Physiologically, skin conductance responses (SCRs) provided insight into the emotional underpinnings of these decisions. Players who showed higher arousal during anticipatory phases—both when their opponent passed or when they decided to "take"—indicated that emotional responses were influencing their decision-making. Notably, in the *random_rounds* condition, there was a significant difference in SCR between the *known_rounds* and *random_rounds* games, with heightened arousal observed during key decision points. These findings support earlier work in decision-making under risk, such as Van't Wout et al. (2006) in the Ultimatum Game and Bechara et al. (1999) in the Iowa Gambling Task, where emotional arousal was linked to decisions involving loss or gain.

A particularly interesting finding was the negative correlation between risk propensity and SCR during player "take" decisions in the *random_rounds* condition. High-risk players showed lower arousal during self-initiated decisions, which echoes findings from Agren et al. (2019) that high-risk individuals exhibit smaller SCRs during decision-making tasks, regardless of whether the decision is safe or risky. This suggests that SCR may serve as a marker for predisposition to risk-taking, particularly in uncertain scenarios.

In summary, our findings suggest that trust is highly sensitive to uncertainty, with increased risk fostering competitive behavior over cooperation. The physiological data, particularly SCR, support the notion that emotional responses play a critical role in decision-making, especially when uncertainty is involved. These results align with existing literature, highlighting the importance of emotional arousal in guiding decision-making processes, particularly in trust-based and socially-mediated decisions. By integrating behavioral and physiological measures, this study adds to a growing body of evidence showing that emotions—not just rational calculations—are central to strategic decision-making under uncertainty.

Limitations

In this study, both the player and the opponent had the ability to terminate the game, making gameplay outcomes interdependent. As a result, the observed behavior reflects the combined dynamics of two interacting participants rather than isolated individual decision-making. Employing a controlled opponent or algorithmic counterpart in future experiments could help disentangle the specific influence of individual behavioral traits and trustworthiness assessments. Additionally, the current analysis lacks formal methods to distinguish whether trust-based decisions were driven by calculative reasoning (i.e., confidence in expected payoffs) or by affective and emotional biases. Developing analytical approaches to separate these mechanisms would provide deeper insights into the cognitive and emotional foundations of trust behavior.

ANNEXURE

General Trust Survey

Choose one of the ratings on a scale of 1-5 with respect to the statements given below.

1. Most people can be trusted.

    (1- Need to be careful, 5- Most can be trusted surely.)

2. Most people are Fair.

    (1- Most people try to take advantage, 5- Most people try to be fair)

3. Most people are helpful.

(1-Most look out for themselves, 5- Most try to be helpful)

Familiarity Questionnaire

Choose one of the ratings on a scale of 1-5 with respect to the questions given below.

1. Do you know the Opponent?

    (1 - Not at all, 5 - Very well)

2. How long have you known the person?

    (1 - Met just Now, 5- Four or more years)

3. What kind of relationship do you share?

    (1- Stranger 5 - Close Friend)

Trust Score Questionnaire

How much do you agree with the following statement?

1. Most of the time your Opponent would want the best for you.

    (1- Strongly disagree, 5 - Strongly agree)

2. Do you think your Opponent can be trusted?

    (1 - Need to be careful, 5 - S/he can be trusted)

3. Do you think that your Opponent will try to take advantage of you if s/he got the chance or will s/he try to be fair?

    (1 - S/he will try to take advantage, 5 - S/he will be fair)

4. Do you think your Opponent might be helpful to you or s/he will mostly look out for him/herself?

    (1 - S/he will look out for him/herself, 5 - S/he will be helpful

5. Please identify how much you would trust your Opponent for each of the following.

    (1 - Do not trust at all, 5 - Trust Completely)

    1. To repay a loan of 1 lakh rupees.
    2. To keep a secret that is damaging to your reputation.
    3. To provide advice about how best to manage your money.
    4. To look after a child, a family member, or loved one while you are away.

Risk Survey

For each of the following statements, please indicate the likelihood that you would engage in the described activity or behaviour if you were to find yourself in that situation.

Provide a rating on the scale of 1 - 7 (1 - Extremely Unlikely, 7 - Extremely Likely)

1. Admitting that your tastes are different from those of your friends.
2. Investing 10% of your annual income in a start-up based on new technology.
3. Disagreeing with your father on a major issue.
4. Taking a job in a start-up that pays higher but has not got venture capital funding yet.
5. Betting a day's income on the outcome of a high-stakes cricket match.
6. Wearing provocative or unconventional clothes on occasion.
7. Quitting a 25lakh/annum well-settled job to pursue higher studies which require you to take an education loan of 50 lakhs.
8. Taking a job that you enjoy over one that is prestigious but less enjoyable.
9. Taking an extremely difficult new course not taken by any of your friends for the sake of knowledge and learning.
10. Arguing with a friend about an issue on which he or she has a very different opinion.